\documentclass[10pt,journal]{IEEEtran}
\usepackage{cite}
\usepackage[pdftex]{graphicx}
\usepackage[tight,footnotesize]{subfigure}
\graphicspath{{"../PDF/"}}
\usepackage[cmex10]{amsmath}
\usepackage{leftidx}
\usepackage{enumerate}
\usepackage[T1]{fontenc}
\usepackage{multirow}
\usepackage{flushend}
\usepackage{array}
\usepackage[colorlinks=true,urlcolor=blue]{hyperref}
\usepackage[table]{xcolor}
\newcolumntype{L}[1]{>{\raggedright\let\newline\\\arraybackslash\hspace{0pt}}m{#1}}
\newcolumntype{C}[1]{>{\centering\let\newline\\\arraybackslash\hspace{0pt}}m{#1}}
\newcommand{\thickhline}{%
    \noalign {\ifnum 0=`}\fi \hrule height 1pt
    \futurelet \reserved@a \@xhline
}

\usepackage{selinput}
\SelectInputMappings{%
  aacute={á},
  ntilde={ñ},
  Euro={€}
}

\begin{document}
\title{Ultra-high and Tunable Sensitivity Leaky-Wave Scanning Using Gain-Loss C-section Phasers} 
\author{Nghia~Nguyen-Trong,~\IEEEmembership{Graduate~Student~Member,~IEEE,}
	Lianfeng~Zou,~\IEEEmembership{Graduate~Student~Member,~IEEE,}
         Christophe~Fumeaux,~\IEEEmembership{Senior~Member,~IEEE}
	and~Christophe~Caloz,~\IEEEmembership{Fellow,~IEEE,}

\thanks{N. Nguyen-Trong is with School of ITEE, University of Queensland, Australia, 4067 email: n.nguyentrong@uq.edu.au.}
\thanks{L. Zou and C. Caloz are with École Polytechnique de Montréal, 2500, ch. de Polytechnique, H3T 1J4, Montréal, Quesbec, Canada.}
\thanks{ C. Fumeaux is with School of Electrical and Electronic Engineering, University of Adelaide, Australia,}
}
%The paper headers
%\markboth{Transactions on Antennas and Propagation}
\maketitle
%Abstract
\begin{abstract}
A periodic leaky-wave antenna (LWA) with tuning capability and enhanced scanning sensitivity is introduced in this paper. This antenna leverages the concept of active Gain-Loss C-section pairs to tune the group delay of each antenna unit cell without affecting its magnitude response. This in turn changes the scanning angle versus frequency rate in the periodic LWA. The proposed concept is well-suited for the application to real-time spectrum analysis, where it allows the frequency resolution to be tuned or significantly enhanced even with a low-permittivity substrate. 
\end{abstract}

%% Introduction
\section{Introduction} 
\IEEEPARstart{O}{ne} of the fundamental advantages of leaky-wave antennas (LWAs) is their capability of producing highly directive beam with a simple feeding network compared to a phased array. For a typical LWA, the main beam direction $\theta_{\text{m}}$ is calculated as \cite{TWA,olinerLWA}
\begin{equation}
\sin\theta_{\text{m}} = \frac{\beta(\omega)}{k_0} = \frac{\beta(\omega) c_0}{\omega},
\label{eq:thetam}
\end{equation}
where $\beta(\omega)$ is the wavenumber as a function of angular frequency $\omega$, $k_0 = \omega/c_0$ is the free-space wavenumber and $c_0$ is the speed of light in free space. Here $\theta_{\text{m}} = 0$ corresponds to broadside radiation direction. Since LWAs operate in the fast-wave region, the ratio $\beta(\omega)/\omega$ usually varies versus frequency, otherwise the group velocity is greater than the speed of light $c_0$. Therefore, the main beam of an LWA is frequency-dependent. The antenna scanning rate or sensitivity, defined as
\begin{equation}
\zeta(\omega) = \frac{d\theta_m}{d\omega},
\label{eq:zeta}
\end{equation}
measures how fast the main beam scans with frequency.  For point-to-point communications, $\zeta(\omega)$ should be minimized in order for the antenna to maximize the operational bandwidth. Several works have been carried out attempting to mitigate this beam squinting issue \cite{Kiasat2008,Antoniades2008,Shahvarpour2010,Sieven2011,Gomez2016}. On the other hand, the frequency-dependent beam-scanning characteristic of LWAs has been exploited to realize a real-time spectrum analyzer in \cite{Gupta2009}. In this application, a broad-band signal was fed through an LWA. Different frequency components radiate towards different directions and arrive at different instants depending on the signal content. For this type of application, $\zeta(\omega)$ may need to be increased to enhance the frequency scanning resolution. This is conventionally achieved by using a high-permittivity substrate. However, it will be shown later in the paper that $\zeta$ only increases as the square root of relative permittivity, i.e. $\sqrt{\epsilon_r}$, which may require impractically large $\epsilon_\text{r}$. Moreover, the analysis of different types of signals may also require adjustable scanning resolution. Therefore, this paper proposes a technique to enhance and control the scanning sensitivity of LWAs without resorting to high-permittivity substrates. 

The proposed concept is based on a Gain-Loss C-section pair, which has been recently reported in \cite{LF2016}. In this device, adjusting the gain and loss of each C-section tunes the group delay at the resonance frequency while maintaining a constant magnitude response. It will be shown here that when the Gain-Loss C-section pair is integrated in an LWA unit cell, tuning the group delay is equivalent to tuning the scanning sensitivity $\zeta(\omega)$. Based on this principle, an LWA is designed and demonstrated with tunable and enhanced scanning sensitivity.

The paper is organized as follows. Section~\ref{sect:GLPair} reviews the concept of Gain-Loss C-section pair for group delay engineering. The relationship between angle-to-frequency sensitivity $\zeta(\omega)$ and group delay $\tau(\omega)$ is then derived in Sect.~\ref{sect:Principle}. Based on this, Sect.~\ref{sect:UnitCell} describes the design of the antenna unit cell and its dispersion diagram. Finally, Sect.~\ref{sect:Res} demonstrates the scanning resolution enhancement and tunability in an LWA example.

\section{Gain-Loss C-section Phaser}
\label{sect:GLPair}
This section first reviews the concept of the Gain-Loss C-section pair for group delay engineering \cite{LF2016}. The contribution of this paper consists in applying this principle to design a tunable and high-sensitivity LWA, which are demonstrated in Section~\ref{sect:Principle},~\ref{sect:UnitCell} and~\ref{sect:Res}.

\begin{figure}[tbp]
	\centering
	\includegraphics[scale = 1]{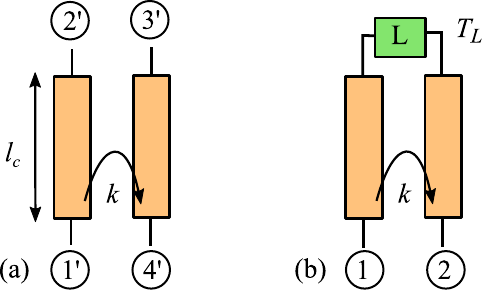}
	\caption{(a) A conventional coupled-line coupler; (b) Loaded C-section with load transfer function $T_L$, created by joining port 2' and 3' of the coupler.}
	\label{fig:Csection}
\end{figure}

\begin{figure}[tbp]
	\centering
	\includegraphics[scale = 0.89]{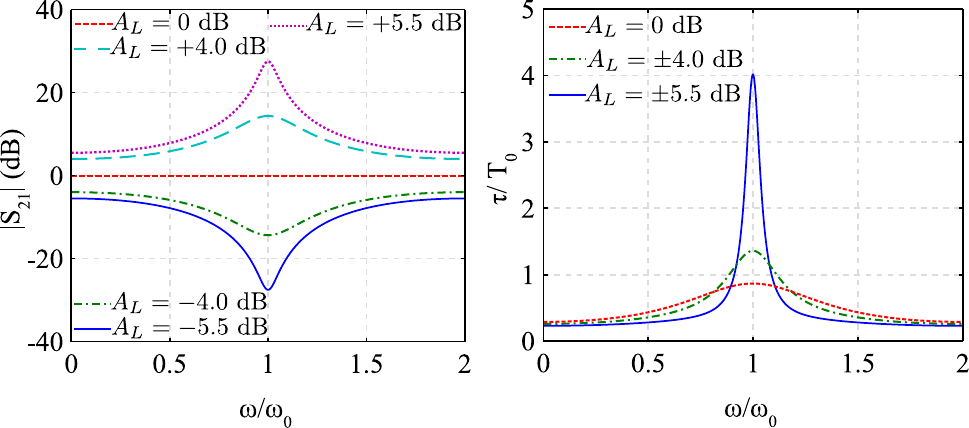}
	\caption{Magnitude (left) and group delay (right) of the C-section with different load transfer functions $T_L = A_L\protect \angle \phi_L$   ($\phi_L = 0$). $T_0 = 1/f_0$ is the period at the resonance frequency $f_0 = \omega_0/2\pi$ of the coupler. The maximum coupling factor is $k = 0.5$ or $k_\text{dB} = -6$~dB.}
	\label{fig:CsectionRes}
\end{figure}

A C-section is obtained by connecting two ends of a coupled line coupler through a load, as shown in Fig.~\ref{fig:Csection}. The load transfer function is denoted $T_L = A_L\angle\phi_L$. Without loss of generality, the phase $\phi_L$ is assumed to be zero, while detailed analysis for arbitrary $\phi_L$ is available in \cite{LF2016}. Under this assumption, the transfer function of the C-section reads \cite{LF2016}
\begin{equation}
S_{21} = A_L\frac{\sqrt{1-k^2} \cot\theta - j(1-k/A_L)}{\sqrt{1-k^2}\cot\theta + j(1-kA_L)},
\label{eq:S21C}
\end{equation}
where $k$ is the maximum coupling occurring at the quarter-wavelength frequency, $\omega_0$, and $\theta = \beta l = (\pi/2) \omega/\omega_0$ is the electrical length of the coupler. The magntiude and phase of the transfer function are derived as

\begin{equation}
|S_{21}| = A_L \sqrt{\frac{(1 - k^2)\cot^2\theta + (1-k/A_L)^2}{ (1 - k^2)\cot^2\theta + (1-kA_L)^2}},
\label{eq:magS21}
\end{equation}
\begin{equation}
\angle S_{21} = -\tan^{-1}\frac{1-k/A_L}{\sqrt{1-k^2}\cot\theta} - \tan^{-1}\frac{1 - kA_L}{\sqrt{1-k^2}\cot\theta}
\label{eq:angleS21}
\end{equation}
Examining~\eqref{eq:magS21} and~\eqref{eq:angleS21} reveals that replacing $A_L$ with $1/A_L$ inverts $|S_{21}|$ is inverted and leaves $\angle S_{21}$ unchanged. According to the second observation, the group delay 
\begin{equation}
\tau = -\frac{d\angle S_{21}}{d\omega},
\label{eq:tau}
\end{equation}
remains unchanged under the transformation $A_L \leftrightarrow 1/A_L$. Figure~\ref{fig:CsectionRes} shows the computed magnitude and group delay responses for different values of $A_L$ at $k = 0.5$ or $k_\text{dB} = -6$~dB. These results suggest the idea of cascading a gain C-section and a loss C-sections, with respective $A_L = G,L$ since in such a system  the gain of the gain C-section \emph{perfectly compensates} for the loss of the loss C-section \emph{at all frequencies} if $G = 1/L$, while the group delay of each C-section pair is simply doubled and may be tuned by varying $G$ and $L$ with $G = 1/L$.
%It can be observed that the phase response is exactly the same when the gain and loss are the same in magnitude (in dB). 

\section{Scanning Sensitivity via Group Delay Tuning in a LWA}
\label{sect:Principle}
The previous section has shown that the Gain-Loss C-section pair can be used to tune the group delay $\tau$ while maintaining a constant magnitude across all frequencies. This result has been utilized to design a distortion-less phaser for real-time signal processing in \cite{LF2016}. In this section, we show that tuning group delay is equivalent to tuning the scanning ratio in a LWA.  

Inserting~\eqref{eq:thetam} into~\eqref{eq:zeta} yields the following relation between the wavenumber and the scanning sensitivity:
\begin{equation}
\zeta(\omega) = \frac{d}{d\omega}\sin^{-1}\left( \frac{\beta c_0}{\omega} \right).
\label{eq:zeta2}
\end{equation}
As the LWA scans across the broadside direction $(\theta_m \approx 0)$, $\beta$ is close to $0$, thus in this range
\begin{eqnarray}
\zeta(\omega) \approx \frac{d}{d\omega} \left( \frac{\beta c_0}{\omega} \right)  &=& c_0 \frac{d}{d\omega}\left( \frac{\beta}{\omega} \right) \nonumber \\
&=& c_0 \left( \frac{1}{\omega}\frac{d\beta}{d\omega} - \frac{\beta}{\omega^2} \right).
\label{eq:zeta3}
\end{eqnarray}

In the case of a periodic LWA, one has $\beta = \phi_\text{u}/p$ where $p$ is the size of the unit cell or period of the LWA, and $\phi_\text{u}$ is the phase shift across the unit cell. The unit-cell group delay is then related to $\beta$ as
\begin{equation}
\tau = \frac{d\phi_u}{d\omega} = p\frac{d\beta}{d\omega}.
\label{eq:tau2}
\end{equation}
Inserting~\eqref{eq:tau2} into~\eqref{eq:zeta3} and considering the broadside region ($\beta\approx 0,\omega\approx\omega_0$), yields
\begin{equation}
\zeta_0 \approx \frac{c_0}{\omega_0 p}\tau_0,
\label{eq:zeta4}
\end{equation}
where $\tau_0$ is the group delay at $\omega_0$. Equation~\eqref{eq:zeta4} shows that the scanning sensitivity is directly related to the group delay, with former being proportional to the latter at broadside. This suggests that the scanning sensitivity can be controlled by tuning the group delay of each unit cell of an LWA, which is achievable in the Gain-Loss C-section pair discussed in Section~\ref{sect:GLPair}.  

\section{LWA Realization}
This section first shows how integrating a Gain-Loss C-section into the LWA unit cell allows one to control the scanning sensitivity.  It next provides design guidelines for the implementaion of the structure. 
\label{sect:UnitCell}
\subsection{Principle}
\begin{figure}[tbp]
	\centering
	\includegraphics[scale = 1]{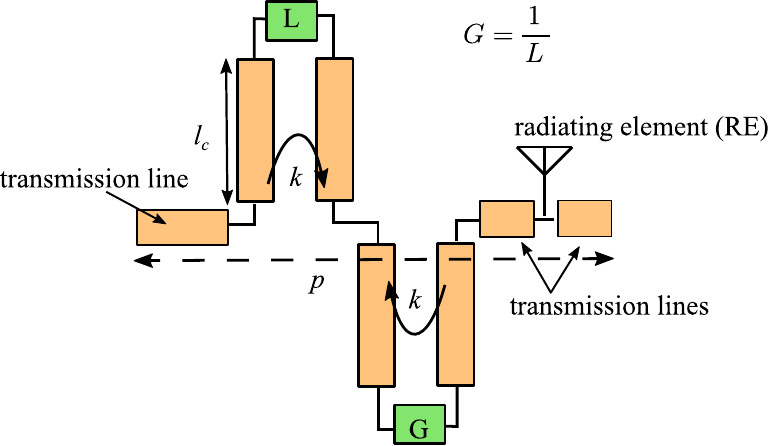}
	\caption{Gain-Loss C-section loaded LWA unit cell.}
	\label{fig:UnitCellPrinciple}
\end{figure}

The conceptual design of the proposed LWA is shown in Fig.~\ref{fig:UnitCellPrinciple}. The unit cell consists of a Gain-Loss C-section pair and a radiating element interconnected by non-dispersive transmission lines. To first understand the operational principle, we assume the radiating element is perfectly matched, non-dispersive with zero phase delay for simplicity. We also assume the transmission lines and the couplers are lossless. Since the gain and loss from the C-section pair perfectly compensate for each other, under the above assumption, the only loss from the structure is from the radiation of the element RE (Fig.~\ref{fig:UnitCellPrinciple}). The leakage factor is then calculated as
\begin{equation}
\alpha(\omega) = \frac{1}{p}\ln |S_{21,\text{UC}}(\omega)| = \frac{1}{p}\ln |S_{21,\text{RE}}(\omega)|
\label{eq:alpha}
\end{equation}
where $S_{21,U}$ and $S_{21,RE}$ are the transmission coefficient of the unit-cell and the radiating element, respectively. Since constant $\alpha(\omega)$ is required for stable gain across scanning frequency \cite{Otto2014}, \eqref{eq:alpha} shows that a constant $|S_{21,RE}(\omega)|$ is desirable.

The results in Fig.~\ref{fig:CsectionRes} show that the Gain-Loss C-section pair can tune the group delay, hence the scanning sensitivity (see Section~\ref{sect:Principle}), at the resonance frequency $\omega_0$ of the coupler.  In the present example, $\omega_0$ is chosen as the broadside frequency so that one can tune the scanning sensitivity at broadside. It is noted that when operating at $\omega_0$, the phase shift of each C-section is $\pi$, i.e. corresponding to two quarter-wave length sections: $l_c = \lambda_0/4$. Therefore, the total phase shift of a Gain-Loss C-section pair is $2\pi$. For $\omega_0$ to be the broadside frequency, the total phase shift within a unit-cell needs to be a multiple of $2\pi$, i.e. $2\pi m_1 \equiv 0$ where $m_1$ is an integer. In the present case, the transmission lines in Fig.~\ref{fig:UnitCellPrinciple} is designed such that its total phase shift is $2\pi$. Hence, the total phase shift of the unit-cell at $\omega_0$ is $2\pi + 2\pi = 4\pi$.

\begin{figure}[tbp]
	\centering
	\includegraphics[scale = 1]{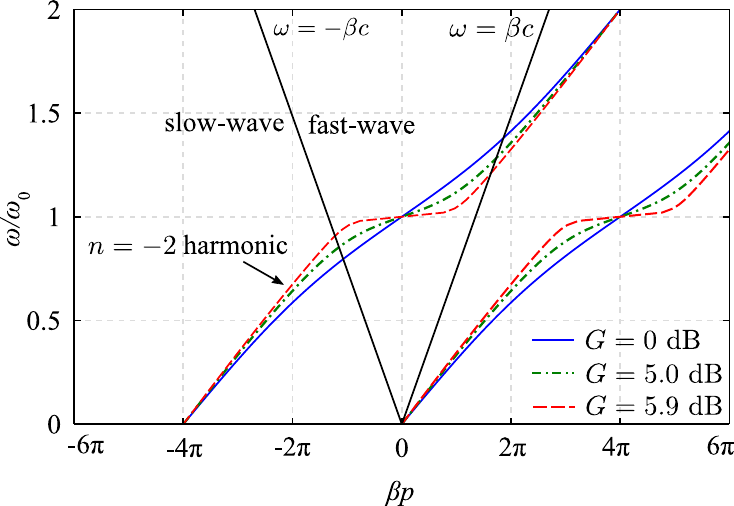}
	\caption{Dispersion diagram of the LWA in Fig.~\ref{fig:UnitCellPrinciple} for $k = 0.5$ or $k_\text{dB} = -6$~dB, calculated using~\eqref{eq:angleS21}.}
	\label{fig:Disp}
\end{figure}
Figure~\ref{fig:Disp}  shows the dispersion diagram of the conceptual unit-cell shown in Fig.~\ref{fig:UnitCellPrinciple}. The maximum coupling factor $k$ is chosen as $k = 0.5 = -6$~dB and can be adjusted when designing the couplers. As explained above, the phase delay of the whole unit-cell is $4\pi$ at broadside frequency $\omega_0$. Therefore, this periodic LWA operates in the $n = -2$ harmonics. It can be verified that by increasing the gain and loss of the C-sections, the scanning sensitivity can be improved. 

Inserting~\eqref{eq:angleS21} into \eqref{eq:tau} and evaluating the group delay at $\omega = \omega_0$ yields 
\begin{eqnarray}
\tau_0 &=& -\frac{d\angle S_{21}}{d\omega}|_{\omega = \omega_0} \\ \nonumber
&=& \frac{T_0\sqrt{1-k^2}}{4}\left( \frac{1}{1- k/A_L} + \frac{1}{1-kA_L}\right).
\end{eqnarray}
Therefore, at the limit $A_L = G \rightarrow 1/k^-$, the group delay reaches
\begin{eqnarray}
\lim_{A_L \rightarrow 1/k^-} \tau_0 = +\infty.
\end{eqnarray}
Since at broadside the scanning sensitivity is proportional to the group delay (Sect.~\ref{sect:Principle}), $\zeta$ also reaches infinity at this limit (the dispersion curve in Fig.~\ref{fig:Disp} is flat at $\omega_0$). This has been shown in \cite{LF2016} as the stability limit of the Gain-Loss C-section pair. Increasing $G$ further will cause the power trapped within the couplers go to infinity, which will obviously damage the device \cite{}. Nevertheless, theoretically, the antenna scanning sensitivity $\zeta$ can be tuned up to infinity (but not at infinity) while keeping the device within the stability region. 

\subsection{Practical design}
%- Targets: to increase the scanning sensitivity for scanning angle from $-45^{\circ}$ to $45^{\circ}$ with no grating lobe ($\theta = 0^{\circ}$ means broadside) 
%A practical design of the unit-cell of SFP LWA is shown in Fig.~\ref{fig:SFP}. In this design, the load is a variable-gain amplifier with transitioning microstrip lines. 

Based on the concept presented above, a unit-cell design using a series-fed microstrip patch (SFP) LWA is demonstrated in this section. This type of antenna is chosen due to its simple geometry and ease of fabrication. For a practical design, several implementation issues need to be solved, which are discussed as below. 

First, the two loads of the Gain-Loss C-section pair need to be realized as active variable-gain amplifiers, in order to provide tunable gain and loss in each C-section. This device is inherently non-reciprocal \cite{Taravati2016}, which makes the design procedure different from a conventional periodic LWA. Typically, there is no need to obtain wideband matching for an LWA unit-cell, i.e $|S_{11,U}|$ is not necessary zero across the scanning frequency range. This is because when the unit-cells are cascaded, the multiple reflections along the antenna will destructively interfere with each other, resulting in satisfactory reflection coefficient as long as the whole antenna is matched with its Bloch impedance. The issue in the conventional LWA design is, on the other hand, at the broadside frequency where there are constructive interferences of multiple reflections, which results in higher reflection coefficient at broadside or even a bandgap.  In the current design, since the amplifiers are non reciprocal with typical parameter $S_{12,A} = 0$, multiple reflections along the antenna are significantly absorbed into these devices. Therefore, each unit-cell \textit{needs to be matched in the scanning frequency range}, otherwise the antenna efficiencies at frequencies $\omega \neq \omega_0$ are significantly reduced which ultimately results in unbalancing gain. If this is achieved, the issue at broadside in the conventional LWA is automatically solved because the reflections along the antenna, if there is any, are significantly absorbed into the amplifiers ($S_{12,A} \approx 0, S_{22,A} \approx 0$).
 
Second, due to the finite size of the amplifier, connecting the chip amplifier with port 2' and 3' of the coupler (Fig.~\ref{fig:Csection}a) requires a transmission line. In order to provide zero phase for the load $\phi_L = 0$, the total phase delay of this transmission line and the amplifier should be $2m_2\pi$ where $m_2$ is an integer. This can be achieved simply by designing and calibrating the amplifier separately before adding into the whole structure \cite{LF2016}. 

\begin{figure}[tbp]
	\centering
	\includegraphics[scale = 0.92]{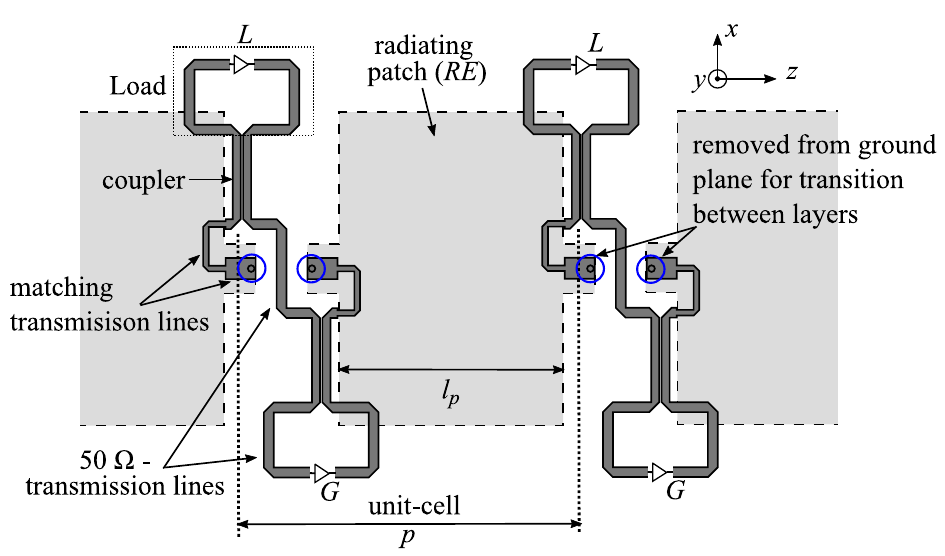}
	\caption{A practical design of a SFP unit-cell for tunable scanning sensitivity application (bottom view). The unit-cell length is $p = 30.5~\text{mm} = 0.51\lambda_0$ where $\lambda_0$ is the free-space wavelength at $\omega_0$. Darker shaded area is copper in the bottom layer. Lighter shaded area bounded by dashed line is copper in the top layer. The top and bottom layer are connected through vias with the ground plane in the middle.}
	\label{fig:SFPUC}
\end{figure}

Based on the above discussion, a practical design for a unit-cell SFP antenna is shown in Fig.~\ref{fig:SFPUC}. A two-layer structure is utilized to isolate the radiating elements $RE$, i.e. microstrip patch on top, with the Gain-Loss C-section pair at bottom. The radiating patch is wideband matched with two quarter-wave length microstrip line sections. The amplifier is of type MAAM-011100 (MACOM Technical Solutions) with input and output impedance of 50~$\Omega$. It is noted that the loads in Fig.~\ref{fig:SFPUC} include the amplifer and its transitioning section with length chosen such that the phase $\phi_L = 2\pi \equiv 0$. 

In the configuration proposed in Fig.~\ref{fig:SFPUC}, the microstrip transmission lines can be oriented such that the unit-cell length $p$ can be chosen arbitrarily (as long as it is larger than the patch length).  This is a design advantage because a smaller unit-cell length can avoid grating lobe \cite{Gupta2014}. In the present case, the length of the unit-cell is chosen as $p = 30.5$~mm such that there is no grating lobe when scanning from $\theta = -30^{\circ}$ to~$30^{\circ}$.

\begin{figure}[tbp]
	\centering
	\includegraphics[scale = 1]{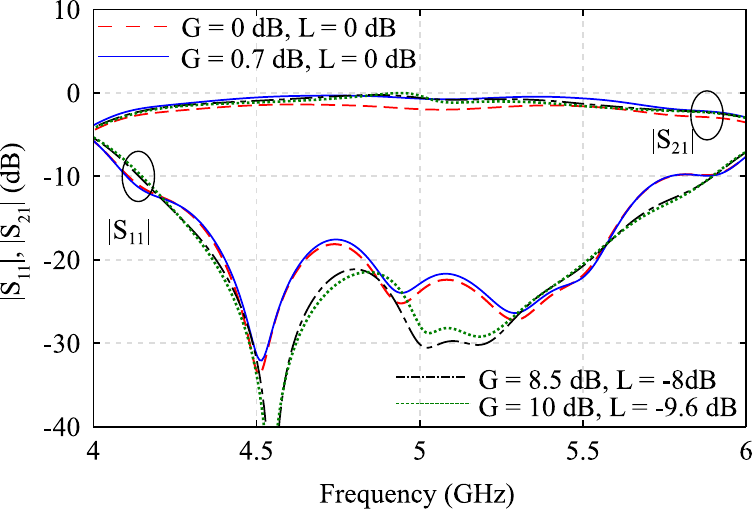}
	\caption{Simulated S-parameter of the optimized unit-cell for different load transfer function values $G$ and $L$.}
	\label{fig:S1}
\end{figure}

The unit-cell is designed and simulated in CST for $f_0 = 5$~GHz. The top substrate, i.e. for radiating patches, is chosen as Roger Duroid 5880 with relative permittivity $\epsilon_{r1} = 2.2$, thickness $h_1 =  0.787$~mm and loss tangent $\tan\delta _1= 0.0009$. The bottom substrate, i.e. for Gain-Loss C-section pairs, is chosen as Roger Duroid 6006 with relative permittivity $\epsilon_{r2} = 6.15$, thickness $h_2 = 0.635$~mm and loss tangent $\tan\delta_2 = 0.0027$. Discrete ports are used to model the variable-gain amplifiers. The amplifiers are assumed to be ideal with input and output impedance of $50~\Omega$, $S_{12,A} = 0$ and $S_{21,A} = G, L$. For simple design, the couplers are realized as coupled microstrip lines as shown in Fig.~\ref{fig:SFPUC}. The coupler should be designed separately to obtain satisfactory matching and isolation in the scanning frequency range around $f_0 = 5$~GHz before being added into the whole unit-cell. The maximum coupling factor is designed to be $k = -10.5$~dB. This value can be adjusted by changing the distance between two coupled microstrip lines. Figure~\ref{fig:S1} shows the simulated S-parameters of the optimized unit-cell for different combinations of $G$ and $L$. To obtain these results, the unit-cell is matched with $50~\Omega$ waveports. 

In Fig.~\ref{fig:S1}, when $G = L = 0$~dB, i.e. no gain or loss in the loads, the unit-cell transmission coefficient is approximately $-2$~dB around resonance frequency $f_0 = 5$~GHz. This loss includes the dielectric, conductor and radiation loss of the unit-cell. These losses can be compensated by choosing the gain $G$ slightly higher than the loss $L$ (in dB-magnitude). This approach has been justified in \cite{Gupta2016amplitude}. It can be observed from Fig.~\ref{fig:S1} that when $G$ is chosen to be slightly higher than $L$ (in dB-magnitude), the transmission coefficient is almost $0$~dB in a wide range of frequency around $f_0$ (due to wideband matching design in Fig.~\ref{fig:SFPUC}). The amplification to compensate for the loss in each unit-cell (or group of unit-cells) improves the directivity of the LWA as demonstrated in \cite{Miranda2006}, which is also an advantage of the present design. 

\begin{figure}[tbp]
	\centering
	\includegraphics[scale = 1]{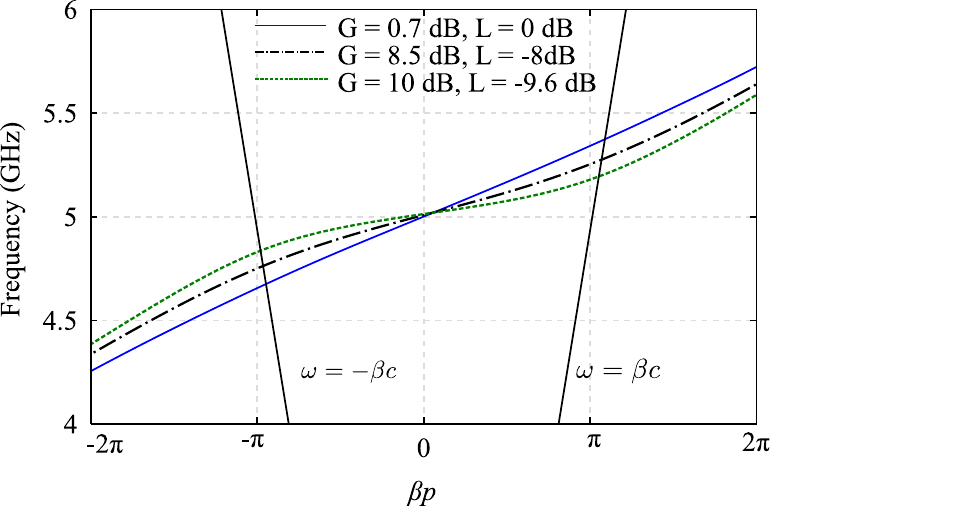}
	\caption{Simulated dispersion diagram of the optimized unit-cell for different load transfer function values $G$ and $L$.}
	\label{fig:DispSim}
\end{figure}

Due to the additional quarter-wavelength matching sections and the transitions for the amplifiers, the total phase delay of a unit-cell is $10\pi$, noting that for a typical SFP LWA this value is $2\pi$. This means that the antenna operates in $n = -5$-harmonic. Figure~\ref{fig:DispSim} shows the simulated dispersion diagram of the unit-cell proposed in Fig.~\ref{fig:SFPUC} for different combinations of $(G,L)$. In this design, the scanning sensitivity, i.e. $\zeta \propto d\beta/d\omega$, is enhanced by two factors:
\begin{enumerate}
\item The electrical length of each unit-cell is increased with additional transmission lines while the physical length $p$ is fixed at a constant value (such that there is no grating lobe). Since it takes more time for power to travel through one unit-cell, this is equivalent to increasing the permittivity of the substrate and thus also enhancing the scanning sensitivity as shown later in Section~\ref{sect:SenEn}. 

\item The group delay in each unit-cell is increased by increasing the gain and loss in the C-section pair. With the control of Gain-Loss C-section pair, $\zeta$ not only can be increased to an arbitrarily large value while keeping the C-section within its stability region, but also is tunable in a certain range. 
\end{enumerate}

\section{Results}
\label{sect:Res}
This section shows numerical results for the SFP LWA designed as shown in Fig.~\ref{fig:SFPUC}. Five unit-cells are cascaded ($N = 5$) and terminated by another radiating patch. Thus the structure consists of 5 Gain-Loss C-section pairs and 6 radiating elements.  The antenna is matched with 50~$\Omega$ coaxial connector at both terminals. The total antenna length is $l_{\textrm{ant}} = Np + l_p = 171.9~\textrm{ mm} \approx 2.87\lambda_0$  where $l_p$ is the length of the patch and $\lambda_0$ is the free-space wavelength at $f_0 = 5$~GHz.
\subsection{S-Parameters}
\label{sect:S}
\begin{figure}[tbp]
	\centering
	\includegraphics[scale = 1]{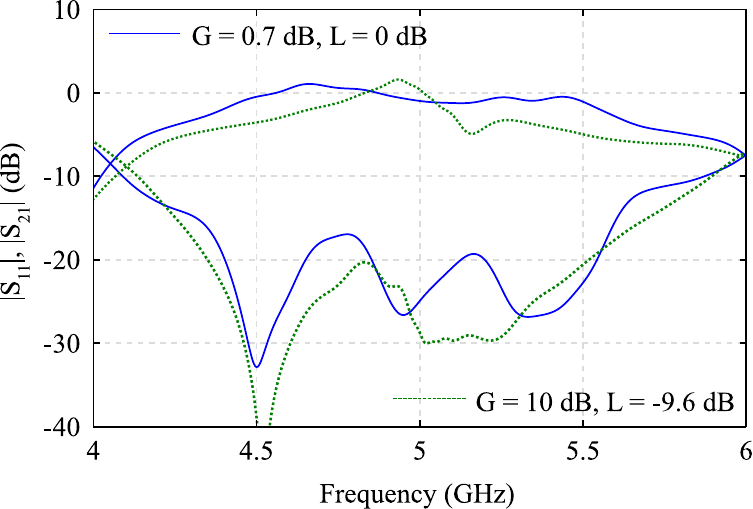}
	\caption{Simulated S-parameter of the antenna for different load transfer function values $G$ and $L$.}
	\label{fig:S2}
\end{figure}

The S-parameters of the antenna for different values of $(G,L)$ are shown in Fig.~\ref{fig:S2}. It is noted that $(G,L)$ are chosen such that the transmission coefficient of each unit cell $|S_{21}|$  is 0~dB around the broadside frequency $\omega_0$ (Fig.~\ref{fig:S1}). It can be observed that by cascading $N = 5$ unit-cells, the overall $|S_{21}|$ varies more significantly in the scanning frequency range (around $5$~GHz), especially at high values of $G$ and $L$. This can be explained as follows. From~Fig.~\ref{fig:CsectionRes}, one can conclude that for the Gain C-section to perfectly compensate the loss from the Loss C-section, the coupler in each section must be identical such that their resonance frequency is exactly the same. However, in practice, these resonance frequency might be slightly shifted due to the distributed loss in the transmission lines, the imperfect isolation and matching of the couplers and coupling effects from other elements. Such a small shift can become significance when $G$ and $L$ increases (see Fig.~\ref{fig:S1} for the case of $G = 10$~dB and $L = -9.6$~dB). Finally, cascading multiple unit-cells causes the unbalance in $|S_{21}|(\omega)$ adding up as shown in Fig.~\ref{fig:S2}. This unbalance in $|S_{21}|(\omega)$ of each unit-cell affects the gain the scanning frequency range as will be shown in the next section. 

\subsection{Scanning Sensitivity Enhancement}
\label{sect:SenEn}
To demonstrate the scanning sensitivity enhancement, we will compare the scanning sensitivity of our design with a passive SFP LWA without Gain-Loss C-section pairs. This classical SFP LWA is designed in the same substrate, i.e. Rogers Duroid 5880 with $\epsilon_{r1} = 2.2$ and $h_1 = 0.787$~mm with five unit cells. The length of each unit-cell is 
\begin{equation}
l^* = \lambda_g = \frac{2\pi c_0}{\omega_0\sqrt{\epsilon_{\textrm{eff},r1}}}
\label{eq:ls}
\end{equation}
where $\lambda_g$ is the guided wavelength in the substrate at broadside frequency $\omega_0$ and $\epsilon_{\textrm{eff},r1}$ is the effective relative dielectric constant of the substrate (noting the the superscript * is used for the passive SFP LWA). This makes the phase delay in each unit-cell $2\pi$ at $f_0$. The total length of this antenna is $170.5$~mm which is compatible to the design for scanning-sensitivity enhancement. Assuming that the radiating patch is non-dispersive, the phase constant is $\beta^* = 2\pi/\lambda_g$. Substitute this and~\eqref{eq:ls} into~\eqref{eq:zeta3}, the angle-to-frequency ratio at around broadside frequency $\omega_0$ of the passive SFP LWA is estimated as
\begin{equation}
\zeta_0^* \approx \frac{\sqrt{\epsilon_{\textrm{eff,r1}}}}{\omega_0},
\label{eq:zetas}
\end{equation}
In practice, due to the dispersion of the radiating patch, the true value of $\zeta_0^*$ is usually higher (but still in the order of~\eqref{eq:zetas}). Nevertheless, this shows that the scanning sensitivity is proportional to the square root of effective relative permittivity of the substrate. 

\begin{figure}[tbp]
	\centering
	\includegraphics[scale = 0.95]{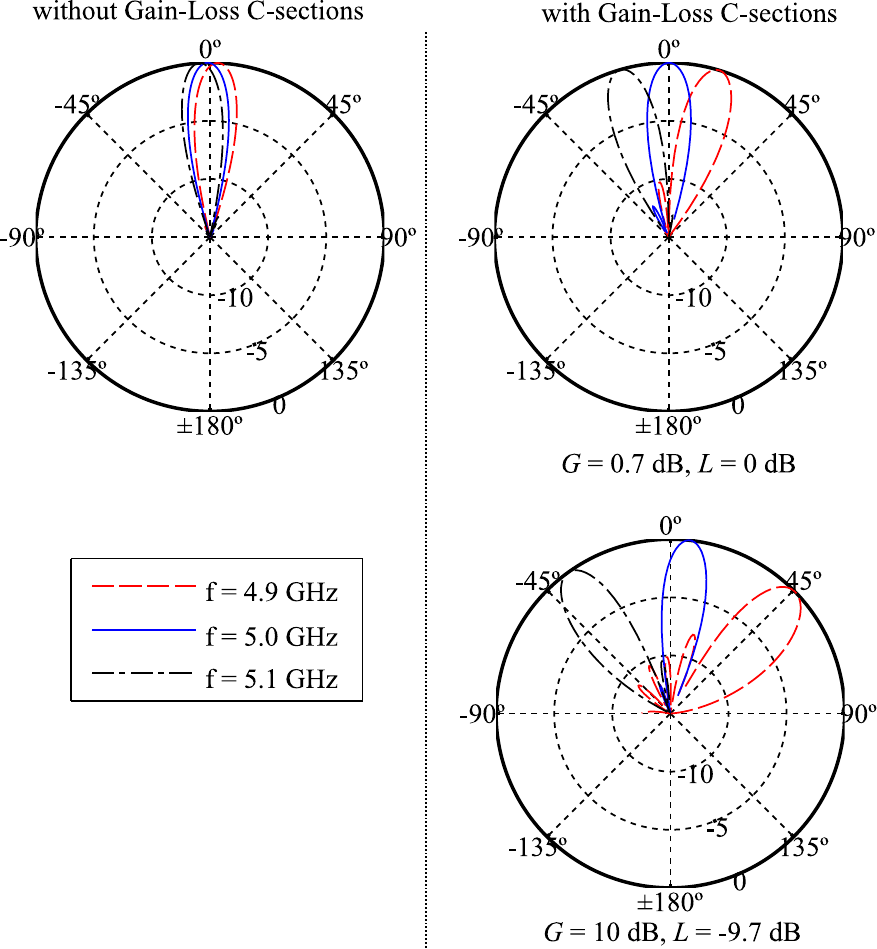}
	\caption{Simulated radiation patterns of the passive SFP LWA (left) and active Gain-Loss C-section SFP LWA (right, shown for two cases of $(G,L)$).}
	\label{fig:Pat}
\end{figure}

Figure~\ref{fig:Pat} shows the simulated normalized radiation patterns of the passive SFP LWA and the active SFP LWA with Gain-Loss C-section pairs. The angle-to-frequency ratio obtained from simulations is estimated as
\begin{equation}
\zeta = \frac{\Delta\theta_m}{\Delta \omega} \approx \frac{\theta_{m1} - \theta_{m2}}{\omega_1 - \omega_2},
\end{equation}
where $\theta_{m1}$ and $\theta_{m2}$ are the main beam direction of two close frequencies $f_1$ and $f_2$, respectively. The simulated angle-to-frequency ratio at broadside frequency for the passive SFP LWA is $\zeta_0^* = 0.026^{\circ}/\textrm{MHz}$, noting that the calculated value from~\eqref{eq:zetas} is $0.016^{\circ}/\textrm{GHz}$, which is in a reasonable agreement. For the active SFP LWA, the scanning sensitivity is much improved:  $\zeta_0= 0.16^{\circ}/\textrm{MHz}$ when $G = 0.7$~dB, $L = 0$~dB, and $\zeta_0 = 0.42^{\circ}/\textrm{MHz}$ when $G = 10$~dB, $L = -9.6$~dB. This shows that the design can improve the scanning sensitivity by a factor of at least 16 times. It is worth mentioning that to achieve this in a passive SFP LWA, the effective permittivity needs to increase by a factor of $16^2 = 256$. In the frequency range $[4.9, 5.1]$~GHz, the realized gain of the active antenna varies within $[8.7, 10.8]$~dB when $G = 0.7$~dB, $L = 0$~dB and $[7.9, 13.3]$~dB when $G = 10$~dB, $L = -9.6$~dB. The unbalance in realized gain of the antenna at higher values of $(G,L)$ is due to the unbalance of $|S_{21}(\omega)|$ across frequency of each unit-cell as shown in Figs.~\ref{fig:S1} and~\ref{fig:S2}.

\subsection{Further Discussion}
For the above design, theoretically, the scanning sensitivity can be tuned with angle-to-frequency ratio from $\zeta_0= 0.16^{\circ}/\textrm{MHz}$ up to an arbitrarily large value when the gain $G$ approaches $1/k$, i.e. $G \rightarrow 10.5$~dB. At this limit, the group delay at the resonance frequency $\omega_0$ is infinity and $|S_{21}|$ of the Gain C-section becomes infinity which damages the device. Furthermore, scanning with a very high resolution also consumes a large amount of power as $|S_{21}|$ of the Gain C-section becomes very large. 

Finally, the enhancement of scanning sensitivity is also limited by fabrication tolerance. This is because due to fabrication tolerances, the resonance frequency of each C-section can be slightly shifted. As discussed in Section~\ref{sect:S}, a small shift in resonance frequency can cause highly unbalance $|S_{21}(\omega)|$ in the antenna especially at the larger values of $G$ and $L$. This in turns causes the unbalance in the antenna gain. 
 
\section{Conclusion}
A LWA that has the capability of tuning and enhancing the scanning sensitivity has been proposed. The design utilizes Gain-Loss C-section phasers to tune the  group delay in each unit-cell while maintaining a constant transmission coefficient. Theoretically, the scanning sensitivity can be increased to an arbitrarily large value while remaining in the stable region of the Gain C-section. In practice, there is a trade-off between the scanning sensitivity, power consumption and fabrication tolerances. This antenna is one of the best candidates for the real-time spectrum analysis application, where the scanning sensitivity is required to adjusted and improved.

\bibliographystyle{IEEEtran}
\bibliography{IEEEabrv,HighSensitivityLWA}
\end{document}